\documentclass{article}
\usepackage{cite}
\usepackage{graphicx}
\usepackage{dcolumn}

\begin{document}

\date{}
\title{Algebraic analysis of non-Hermitian quadratic Hamiltonians}
\author{Francisco M. Fern\'{a}ndez\thanks{%
fernande@quimica.unlp.edu.ar} \\
INIFTA, DQT, Sucursal 4, C. C. 16, \\
1900 La Plata, Argentina}
\maketitle

\begin{abstract}
We study a general one-mode non-Hermitian quadratic Hamiltonian that does
not exhibit $\mathcal{PT}$-symmetry. By means of an algebraic method we
determine the conditions for the existence of real eigenvalues as well as
the location of the exceptional points. We also put forward an algebraic
alternative to the generalized Bogoliubov transformation that enables one to
convert the quadratic operator into a simpler form in terms of the original
creation and annihilation operators. We carry out a similar analysis of a
two-mode oscillator that consists of two identical one-mode oscillators
coupled by a quadratic term.
\end{abstract}

\section{Introduction}

\label{sec:intro}

Some time ago, Swanson\cite{S04} studied a non-Hermitian quadratic
oscillator and obtained the energy eigenstates by means of the generalized
Bogoliubov transformation of the creation and annihilation operators. Since
the eigenstates exhibit imaginary norm he found the orthonormal dual space
that allows the inner product to be redefined using a complexification
procedure.

A simple algebraic method based on ladder operators proved useful for the
treatment of a variety of quadratic Hamiltonians, both Hermitian and
non-Hermitian, in terms of coordinates and momenta\cite
{F15a,F15b,F16a,F16b,F20}. The reason is that many features of a quadratic
Hamiltonian operator can be straightforwardly derived from its regular or
adjoint matrix representation\cite{G74,FC96}. Bagarello\cite{B21} proposed a
most interesting generalization of such abstract ladder operators.

The regular or adjoint matrix representation of the Hamiltonian operator is
closely related to the fundamental matrix that proved to be useful in
determining the conditions under which a $\mathcal{PT}$-symmetric elliptic
quadratic differential operator with real spectrum is similar to a
self-adjoint operator\cite{CGHS12}. This matrix is known since long ago\cite
{H58} and some researchers refer to it as Hoppfield-like\cite{CBC05} or
Hoppfield-Bogoliubov\cite{XXW21} matrix. It proves to be useful for the
treatment of a variety of problems\cite{RK05,RK09,RCR14}.

The purpose of this paper is the application of the algebraic method to the
quadratic non-Hermitian Hamiltonian operator put forward by Swanson\cite{S04}%
. In section~\ref{sec:algebraic} we generalize the algebraic method so that
it may be suitable for a wider range of basis operators and specialize in
the creation and annihilation boson operators. In section~\ref{sec:one-mode}
we apply this approach to the a generalized Swanson oscillator and also put
forward an algebraic variant of the Bogoliubov transformation. In section~%
\ref{sec:two-mode} we study a model given by two identical generalized
Swanson oscillators coupled by a quadratic term. Finally, in section~\ref
{sec:conclusons} we summarize the main results of the paper and draw
conclusions.

\section{Generalized algebraic method}

\label{sec:algebraic}

In earlier papers we have illustrated the application of the algebraic
method to quadratic Hermitian and non-Hermitian Hamiltonians expressed in
terms of coordinate and momenta\cite{F15a,F15b,F16a,F16b,F20} . In many
cases it is more convenient to write such operators in terms of creation and
annihilation boson operators. For this reason, in what follows we develop
the approach in a somewhat more general way that includes both cases.

We focus on a Hamiltonian operator
\begin{equation}
H=\sum_{i=1}^{2K}\sum_{j=1}^{2K}g_{ij}O_{i}O_{j},  \label{eq:H_quadratic}
\end{equation}
that is a quadratic function of the linear operators $O_{i}$. Without loss
of generality we assume that $g_{ij}=g_{ji}$. It is necessary that the
otherwise arbitrary linear operators $O_{i}$ satisfy $\left[ O_{i}\left[
O_{j},O_{k}\right] \right] =0$ for all $i,j,k=1,2,\ldots ,2K$. Under this
condition we define the $2K\times 2K$ matrix $\mathbf{U}$ with elements $%
U_{ij}=\left[ O_{i},O_{j}\right] $ that is skew symmetric: $\mathbf{U}^{t}=-%
\mathbf{U}$, where $t$ stands for transpose.

In is not difficult to verify that
\begin{equation}
\lbrack H,O_{i}]=\sum_{j=1}^{2K}H_{ji}O_{j},  \label{eq:[H,Oi]}
\end{equation}
where the $2K\times 2K$ matrix $\mathbf{H}$ with elements $H_{ij}$ is given
by
\begin{equation}
\mathbf{H}=2\mathbf{GU},  \label{eq:H=2gU}
\end{equation}
where $\mathbf{G}$ is the $2K\times 2K$ symmetric matrix with elements $%
g_{ij}$. $\mathbf{H}$ is commonly called the regular or adjoint matrix
representation of the operator $H$ in the basis $B=\left\{
O_{1},O_{2},\ldots ,O_{2K}\right\} $\cite{F15a,F15b,F16a,F16b,F20,G74,FC96},
closely related to the fundamental matrix\cite{CGHS12} and also called
Hoppfield\cite{H58,CBC05} or Hoppfield-Bogoliubov\cite{XXW21} matrix.

From the Jacobi identity $\left[ H,\left[ O_{i},O_{j}\right] \right] +\left[
O_{j}\left[ H,O_{i}\right] \right] +\left[ O_{i}\left[ O_{j},H\right]
\right] =0$ we conclude that $\left( \mathbf{UH}\right) ^{t}=\mathbf{UH}$ or
$-\mathbf{H}^{t}\mathbf{U}=\mathbf{UH}$. If the set $B$ of $2K$ operators $%
Q_{i}$ is linearly independent, then the determinant of the matrix $\mathbf{U%
}$, $\left| \mathbf{U}\right| $, is nonzero and the inverse $\mathbf{U}^{-1}$
exists. Therefore,
\begin{equation}
\mathbf{UHU}^{-1}=-\mathbf{H}^{t}.  \label{eq:UHU^(-1)=-H^t}
\end{equation}

The algebraic method is based on ladder opetators of the form
\begin{equation}
Z=\sum_{i=1}^{2K}c_{i}O_{i},  \label{eq:Z}
\end{equation}
such that
\begin{equation}
\lbrack H,Z]=\lambda Z.  \label{eq:[H,Z]}
\end{equation}
If $\left| \psi \right\rangle $ is an eigenvector of $H$ with eigenvalue $E$
then $Z\left| \psi \right\rangle $ is eigenvector of $H$ with eigenvalue $%
E+\lambda $:
\begin{equation}
HZ\left| \psi \right\rangle =(E+\lambda )Z\left| \psi \right\rangle .
\label{eq:HZ|Psi>}
\end{equation}
It follows from equations (\ref{eq:[H,Oi]}), (\ref{eq:Z}) and (\ref{eq:[H,Z]}%
) that
\begin{equation}
(\mathbf{H}-\lambda \mathbf{I})\mathbf{C}=0,  \label{eq:HC=-lambdaC}
\end{equation}
where $\mathbf{I}$ is the $2K\times 2K$ identity matrix and $\mathbf{C}$ is
a $2K\times 1$ column matrix with elements $c_{i}$. There are nontrivial
solutions only for those values of $\lambda $ that are roots of the
characteristic polynomial $p(\lambda )=\left| \mathbf{H}-\lambda \mathbf{I}%
\right| $.

According to equation (\ref{eq:UHU^(-1)=-H^t}) we have $p(\lambda )=\left| -%
\mathbf{U}^{-1}\mathbf{H}^{t}\mathbf{U}-\lambda \mathbf{I}\right| =\left|
\mathbf{H}^{t}+\lambda \mathbf{I}\right| =p(-\lambda )$ and conclude that if
$\lambda _{i}$ is a root of $p(\lambda )$ then $-\lambda _{i}$ is also a
root. We arbitrarily label the roots so that $\lambda _{i}=-\lambda
_{2K-i+1} $, $i=1,2,\ldots ,K$. Consequently, if all the roots are real then
they can be arranged as $\lambda _{1}\leq \lambda _{2}\leq \ldots \leq
\lambda _{K}\leq 0\leq \lambda _{K+1}=-\lambda _{K}\leq \lambda
_{K+2}=-\lambda _{K-1}\leq \ldots \leq \lambda _{2K}=-\lambda _{1}$. For
every $\lambda _{i}$ we obtain
\begin{equation}
Z_{i}=\sum_{j=1}^{2K}c_{ji}O_{j}.  \label{eq:Z_i}
\end{equation}
It follows from $\left[ H,\left[ Z_{i},Z_{j}\right] \right] =\left( \lambda
_{i}+\lambda _{j}\right) \left[ Z_{i},Z_{j}\right] =0$ that $\left[
Z_{i},Z_{j}\right] =0$ if $\lambda _{i}+\lambda _{j}\neq 0$.

If we normalize the ladder operators as
\begin{equation}
\left[ Z_{i},Z_{2K-i+1}\right] =1,\;i=1,2,\ldots ,K,  \label{eq:Z_i_norm}
\end{equation}
then we can write the Hamiltonian operator as
\begin{equation}
H=\frac{1}{2}\sum_{i=1}^{K}\lambda _{2K-i+1}\left(
Z_{i}Z_{2K-i+1}+Z_{2K-i+1}Z_{i}\right) .  \label{eq:H(Z_i)}
\end{equation}

In earlier papers we considered operators that are quadratic functions of
the coordinate and momenta\cite{F15a,F15b,F16a,F16b,F20}. In what follows,
we specialize in the boson creation $a_{i}^{\dagger }$ and annihilation $%
a_{i}$ operators, $i=1,2,\ldots ,K$, that satisfy the commutation relations $%
\left[ a_{i},a_{j}\right] =0$, $\left[ a_{i}^{\dagger },a_{j}^{\dagger
}\right] =0$ and $\left[ a_{i},a_{j}^{\dagger }\right] =\delta _{ij}$. If we
choose the operator basis set $\left\{ O_{1},O_{2},\ldots ,O_{2K}\right\}
=\left\{ a_{1},a_{2},\ldots ,a_{K},a_{1}^{\dagger },a_{2}^{\dagger },\ldots
,a_{K}^{\dagger }\right\} $ then the matrix $\mathbf{U}$ becomes
\begin{equation}
\mathbf{U}=\left(
\begin{array}{ll}
\mathbf{0}_{K} & \mathbf{I}_{K} \\
-\mathbf{I}_{K} & \mathbf{0}_{K}
\end{array}
\right) ,  \label{eq:U_matrix}
\end{equation}
where $\mathbf{0}_{K}$ and $\mathbf{I}_{K}$ are the $K\times K$ zero and
identity matrices, respectively. The matrix $\mathbf{U}$ is orthogonal and $%
\mathbf{UU}=-\mathbf{I}$.

\section{The one-mode model}

\label{sec:one-mode}

Swanson\cite{S04} studied the quadratic Hamiltonian $H=\omega \left(
a^{\dagger }a+\frac{1}{2}\right) +\alpha a^{2}+\beta a^{\dagger 2}$, where $%
\omega >0$ and $\alpha $, $\beta $ are real model parameters. It is a
particular case of the quadratic Hamiltonian (\ref{eq:H_quadratic}) with $%
K=1 $ and $\left\{ O_{1},O_{2}\right\} =\left\{ a,a^{\dagger }\right\} $.
Since $\mathcal{PT}:\left( a,a^{\dagger }\right) \rightarrow \left(
-a,-a^{\dagger }\right) $, where $\mathcal{P}$ and $\mathcal{T}$ are the
parity and time-reversal operators, respectively, then $H$ is $\mathcal{PT}$
symmetric: $\mathcal{PT}H\mathcal{TP}=H$. Swanson showed that this
Hamiltonian exhibits real eigenvalues provided that $\alpha \beta <\omega
^{2}/4$.

On taking into account that $H(\omega ,\alpha ,\beta )=\omega
^{-1}H(1,\alpha /\omega ,\beta /\omega )$ then we can set $\omega =1$
without loss of generality and in what follows we consider the two-parameter
model
\begin{equation}
H=\frac{1}{2}\left( a^{\dagger }a+aa^{\dagger }\right) +\alpha a^{2}+\beta
a^{\dagger 2}.  \label{eq:H:Swanson_1D}
\end{equation}
In order to make our analysis somewhat more general, we will assume that $%
\alpha $ and $\beta $ may be complex model parameters. The resulting
operator is not $\mathcal{PT}$-symmetric because $\mathcal{PT}H(\alpha
,\beta )\mathcal{TP}=H\left( \alpha ^{*},\beta ^{*}\right) $.

The regular matrix representation of the quadratic Hamiltonian (\ref
{eq:H:Swanson_1D}) reads
\begin{equation}
\mathbf{H}=\left(
\begin{array}{cc}
-1 & 2\alpha \\
-2\beta & 1
\end{array}
\right) .  \label{eq:reg_mat_1D}
\end{equation}
One can easily obtain its eigenvalues
\begin{equation}
\lambda _{1}=-\lambda _{2}=-\sqrt{1-4\alpha \beta },
\label{eq:lambda_(1,2)_1D}
\end{equation}
and corresponding ladder operators
\begin{eqnarray}
Z_{1} &=&N_{1}\left( a+\frac{1-\sqrt{1-4\alpha \beta }}{2\alpha }a^{\dagger
}\right) ,  \nonumber \\
Z_{2} &=&N_{2}\left( a+\frac{1+\sqrt{1-4\alpha \beta }}{2\alpha }a^{\dagger
}\right) .  \label{eq:Z_(1,2)_1D}
\end{eqnarray}
From the commutation condition $\left[ Z_{1},Z_{2}\right] =1$ we obtain $%
N_{1}N_{2}=\frac{\alpha }{\sqrt{1-4\alpha \beta }}$, so that
\begin{equation}
H=\sqrt{1-4\alpha \beta }\left( Z_{2}Z_{1}+\frac{1}{2}\right) ,
\label{eq:H_1DD_Z2Z1}
\end{equation}
according to equation (\ref{eq:H(Z_i)}). There are real eigenvalues provided
that $1-4\alpha \beta >0$, and this result is valid for both the Hermitian ($%
\alpha =\beta ^{*}$) and non-Hermitian ($\alpha \neq \beta ^{*}$) cases.
When $\alpha \beta =1/4$ we have only one ladder operator which indicates
that we are in the presence of an exceptional point (EP). More precisely,
when $\beta =1/(4\alpha )$ the characteristic polynomial of $\mathbf{H}$ is $%
p(\lambda )=\lambda ^{2}$ but there is only one linearly-independent
eigenvector with eigenvalue $\lambda =0$. In other words: the matrix $%
\mathbf{H}$ is not diagonalizable (defective) at the EP. It is worth
pointing out that an EP in $\lambda $ is also an EP in the eigenvalues of $H$%
, as follows from equation (\ref{eq:HZ|Psi>}).

We can proceed in a different way by considering an invertible operator $%
S=e^{Q}$. If $Q$ is a quadratic function of $a$ and $a^{\dagger }$ then $%
SaS^{-1}=s_{11}a+s_{12}a^{\dagger }$ and $Sa^{\dagger
}S^{-1}=s_{21}a+s_{22}a^{\dagger }$\cite{FC96}. It is clear that the
operators $H$ and $\tilde{H}=SHS^{-1}$ are isospectral and that $\left[
SaS^{-1},Sa^{\dagger }S^{-1}\right] =S\left[ a,a^{\dagger }\right] S^{-1}=1$%
. If we require that $\left[ a^{\dagger },\left[ a^{\dagger },\tilde{H}%
\right] \right] =0$ and $\left[ a,\left[ a,\tilde{H}\right] \right] =0$ then
we derive two additional equations for the coefficients $s_{ij}$. Therefore,
from the three conditions just mentioned
\begin{eqnarray}
s_{11}s_{22}-s_{12}s_{21}-1 &=&0,  \nonumber \\
\alpha s_{11}^{2}+\beta s_{21}^{2}+s_{11}s_{21} &=&0,  \nonumber \\
\alpha s_{12}^{2}+\beta s_{22}^{2}+s_{12}s_{22} &=&0,
\label{eq:c_(ij)_conditions}
\end{eqnarray}
we can obtain three coefficients $s_{ij}$ in terms of the remaining one; for
example,
\begin{eqnarray}
s_{12} &=&-\frac{\beta }{\left| s_{11}\right| \sqrt{1-4\alpha \beta }},
\nonumber \\
s_{21} &=&\frac{\left| s_{11}\right| \sqrt{1-4\alpha \beta }}{2\beta }-\frac{%
s_{11}}{2\beta },  \nonumber \\
s_{22} &=&\frac{1}{2\left| s_{11}\right| \sqrt{1-4\alpha \beta }}+\frac{1}{%
2s_{11}}.  \label{eq:c_(ij)}
\end{eqnarray}
If we choose the arbitrary coefficient $s_{11}$ to be real then
\begin{equation}
\tilde{H}=\sqrt{1-4\alpha \beta }\left( a^{\dagger }a+\frac{1}{2}\right) .
\label{eq:UHU}
\end{equation}

If $\mathbf{Q}$ is the regular matrix representation of the quadratic
operator $Q$ and $\mathbf{S}$ the matrix with elements $s_{ij}$ then one can
prove that $\mathbf{S}^{t}=e^{\mathbf{Q}}$\cite{FC96}. From this identity it
is possible to obtain an explicit analytic expression for $Q$. However, it
is not necessary for most purposes\cite{FC96}.

The transformation discussed above maps the eigenvalue equation $H\psi
_{n}=E_{n}\psi _{n}$ into $\tilde{H}\tilde{\psi}_{n}=E_{n}\tilde{\psi}_{n}$,
where $\tilde{\psi}_{n}=S\psi _{n}$. Therefore, $\left\langle \tilde{\psi}%
_{m}\right| \left. \tilde{\psi}_{n}\right\rangle =\left\langle S\psi
_{m}\right| \left. S\psi _{n}\right\rangle =\left\langle \psi _{m}\right|
\left. S^{\dagger }S\psi _{n}\right\rangle =\delta _{mn}$ enables us to
define the Hermitian, positive-definite metric $\rho =S^{\dagger
}S=e^{Q^{\dagger }+Q}$ such that $\left\langle \psi _{m}\right| \left. \psi
_{n}\right\rangle _{\rho }=\left\langle \psi _{m}\right| \left. \rho \psi
_{n}\right\rangle $ as suggested by Pauli\cite{P43} several years ago.

Although this approach resembles the generalized Bogoliubov transformation
used by Swanson\cite{S04}, we believe that it is somewhat more convenient
because the transformed operator $\tilde{H}$ is expressed in terms of the
original boson operators. This canonical transformation, as well as the
generalized Boboliubov one\cite{S04}, breaks down when $\alpha \beta =1/4$
as predicted by the more straightforward algebraic method.

\section{Two-mode model}

\label{sec:two-mode}

In this section, we consider two identical generalized Swanson oscillators
coupled by a quadratic term:
\begin{equation}
H=a_{1}^{\dagger }a_{1}+a_{2}^{\dagger }a_{2}+1+\alpha \left(
a_{1}^{2}+a_{2}^{2}\right) +\beta \left( a_{1}^{\dagger 2}+a_{2}^{\dagger
2}\right) +\gamma \left( a_{1}a_{2}^{\dagger }+a_{1}^{\dagger }a_{2}\right) ,
\label{eq:H_2D}
\end{equation}
where $\alpha $ and $\beta $ are complex and $\gamma $ real.

The regular matrix representation of $H$ is
\begin{equation}
\mathbf{H}=\left(
\begin{array}{llll}
-1 & -\gamma & 2\alpha & 0 \\
-\gamma & -1 & 0 & 2\alpha \\
-2\beta & 0 & 1 & \gamma \\
0 & -2\beta & \gamma & 1
\end{array}
\right) .  \label{eq:regular_mat_2D}
\end{equation}
Its four eigenvalues are given by
\begin{equation}
\lambda _{1}=-\sqrt{\left( \gamma +1\right) ^{2}-4\alpha \beta },\;\lambda
_{2}=-\sqrt{\left( \gamma -1\right) ^{2}-4\alpha \beta },\;\lambda
_{3}=-\lambda _{2},\;\lambda _{4}=-\lambda _{1}.  \label{eq:lambdas_2D}
\end{equation}
In this case the eigenvalues are real provided that $\alpha \beta <\min
\left\{ \left( \gamma +1\right) ^{2},\left( \gamma -1\right) ^{2}\right\} /4$%
. Note that we recover the condition derived in the preceding section when $%
\gamma =0$.

By means of the ladder operators
\begin{eqnarray}
Z_{1} &=&\frac{N_{1}}{2\alpha }\left[ 2\alpha a_{1}+2\alpha a_{2}+\left(
\gamma +1-\sqrt{\left( \gamma +1\right) ^{2}-4\alpha \beta }\right)
a_{1}^{\dagger }+\left( \gamma +1-\sqrt{\left( \gamma +1\right) ^{2}-4\alpha
\beta }\right) a_{2}^{\dagger }\right]   \nonumber \\
Z_{2} &=&\frac{N_{2}}{2\alpha }\left[ 2\alpha a_{1}-2\alpha a_{2}+\left(
1-\gamma -\sqrt{\left( \gamma -1\right) ^{2}-4\alpha \beta }\right)
a_{1}^{\dagger }+\left( \gamma -1+\sqrt{\left( \gamma -1\right) ^{2}-4\alpha
\beta }\right) a_{2}^{\dagger }\right]   \nonumber \\
Z_{3} &=&\frac{N_{3}}{2\alpha }\left[ 2\alpha a_{1}-2\alpha a_{2}+\left(
1-\gamma +\sqrt{\left( \gamma -1\right) ^{2}-4\alpha \beta }\right)
a_{1}^{\dagger }+\left( \gamma -1-\sqrt{\left( \gamma -1\right) ^{2}-4\alpha
\beta }\right) a_{2}^{\dagger }\right]   \nonumber \\
Z_{3} &=&\frac{N_{4}}{2\alpha }\left[ 2\alpha a_{1}+2\alpha a_{2}+\left(
\gamma +1+\sqrt{\left( \gamma +1\right) ^{2}-4\alpha \beta }\right)
a_{1}^{\dagger }+\left( \gamma +1+\sqrt{\left( \gamma +1\right) ^{2}-4\alpha
\beta }\right) a_{2}^{\dagger }\right]   \nonumber \\
N_{1}N_{4} &=&\frac{\alpha }{2\sqrt{\left( \gamma +1\right) ^{2}-4\alpha
\beta }},\;N_{2}N_{3}=\frac{\alpha }{2\sqrt{\left( \gamma -1\right)
^{2}-4\alpha \beta }}  \label{eq:Z_i_2D}
\end{eqnarray}
we rewrite the Hamiltonian operator as
\begin{equation}
H=\frac{1}{2}\left[ \lambda _{4}\left( Z_{1}Z_{4}+Z_{4}Z_{1}\right) +\lambda
_{3}\left( Z_{2}Z_{3}+Z_{3}Z_{2}\right) \right] .  \label{eq:H(Z_i)_2D}
\end{equation}

\section{Conclusions}

\label{sec:conclusons}

In this paper we have studied a more general version of the simple quadratic
Hamiltonian discussed earlier by Swanson\cite{S04} that does not exhibit $%
\mathcal{PT}$ symmetry. We have shown that the algebraic method\cite
{F15a,F15b,F16a,F16b,F20} is suitable for the analysis of this kind of
problems as it reveals the condition for the existence of real eigenvalues
in a simple and clear way. In addition to it, we have put forward an
algebraic alternative version of the generalized Bogoliubov transformation
that enables one to convert the quadratic Hamiltonian into a simpler form in
terms of the original creation and annihilation operators.

We have also studied a model given by two identical generalized Swanson
oscillators coupled by a quadratic term. In its more general form the
resulting two-mode oscillator is neither Hermitian or $\mathcal{PT}$
symmetric but exhibits real eigenvalues under certain conditions. The
analysis is greatly facilitated by the algebraic method that enables one to
obtain the location of the exceptional points in a simple and
straightforward way.

\end{document}